# Gate-tunable Topological Valley Transport in Bilayer Graphene

Mengqiao Sui[1,2], Guorui Chen[1,2], Liguo Ma[1,2], Wenyu Shan[3], Dai Tian[1,2], Kenji Watanabe[4], Takashi Taniguchi[4], Xiaofeng Jin[1,2], Wang Yao[5], Di Xiao[3] and Yuanbo Zhang[1,2*]

[1]*State Key Laboratory of Surface Physics and Department of Physics, Fudan University, Shanghai 200433, China*

[2]*Collaborative Innovation Center of Advanced Microstructures, Shanghai 200433, China*

[3]*Department of Physics, Carnegie Mellon University, Pittsburgh, Pennsylvania 15213, USA*

[4]*Advanced Materials Laboratory, National Institute for Materials Science, 1-1 Namiki, Tsukuba, 305-0044, Japan.*

[5]*Department of Physics and Center of Theoretical and Computational Physics, University of Hong Kong, Hong Kong, China*

*Email: zhyb@fudan.edu.cn

**Valley pseudospin, the quantum degree of freedom characterizing the degenerate valleys in energy bands[1], is a distinct feature of two-dimensional Dirac materials[1–5]. Similar to spin, the valley pseudospin is spanned by a time reversal pair of states, though the two valley pseudospin states transform to each other under spatial inversion. The breaking of inversion symmetry induces various valley-contrasted physical properties; for instance, valley-dependent topological transport is of both scientific and technological interests[2–5]. Bilayer graphene (BLG) is a unique system whose intrinsic inversion symmetry can be controllably broken by a perpendicular electric field, offering a rare possibility for continuously tunable valley-topological transport. Here, we used a perpendicular gate electric field to break the inversion symmetry in BLG, and a giant nonlocal response was observed as a result of the topological transport of the valley pseudospin. We further showed that the valley transport is fully tunable by external gates, and that the nonlocal signal persists up to room temperature and over long distances. These observations challenge contemporary understanding of topological transport in a gapped system, and the robust topological transport may lead to future valleytronic applications.**

In crystalline solids, a topological current can be induced by the Berry phase of the electronic wave function[6]. Examples include the quantum Hall current in a magnetic field, and the spin Hall current arising from spin-orbit coupling. Such topological transport is robust against impurities and defects in materials – a feature that is much sought after in potential electronic applications. In such applications, the ability to switch and to continuously tune the topological transport is crucial. The topological current is in principle dictated by the crystal symmetry, which is difficult to change in bulk materials. Bilayer graphene, however, offer new opportunities in which inversion symmetry can be controllably broken by an external electric field in the perpendicular direction.

The topological current controlled by the inversion symmetry breaking is associated with carriers' valley pseudospin, which characterises the two-fold degenerate band-edges located at the corners of the hexagonal Brillouin zone. The topological Hall current, odd under time-reversal but even under inversion, is strictly zero in pristine mono- and bi-layer graphene which respect both symmetries. When the inversion symmetry is broken, however, time-reversal symmetry requires the Hall currents to have opposite signs and equal magnitudes in the two valleys (*i.e*., a valley Hall effect), as recently demonstrated in monolayer graphene in ref 4. Microscopically, inversion symmetry breaking opens a band gap at the charge neutral point (CNP), and produces valley-contrasted Berry curvatures[2]. The Berry curvatures act similarly to a momentum-space magnetic field that causes the Hall effect at finite doping[6]. In BLG, the inversion symmetry (and the Berry curvature) is for the first time controlled by gate electric field. Valley Hall current is therefore expected to be fully tunable by external gates, bringing topological transport in line with modern electronic technology.

In this study, we demonstrated tunable topological valley Hall transport in gate-biased BLG with nonlocal measurement in a Hall bar configuration (Fig. 1a and Fig. 1c inset), which has been used to detect nonlocal transport in other spin/pseudospin systems[4,7–10]. The charge current injected at one end of Hall bar induced a pure valley Hall current in the transverse direction and, because of the inverse valley Hall effect, converted to a charge imbalance at the other end of the sample; we therefore detected a nonlocal response as a voltage drop. We demonstrated that the giant, tunable, nonlocal response can be induced by the perpendicular gate electric field, but is absent in the gapless BLG in which inversion symmetry is present. Such gate tunability indicates the essential role of inversion symmetry breaking in nonlocal transport, and provides unambiguous evidence that our nonlocal signal was a result of the valley Hall effect. The nonlocal transport persists up to room temperature and over long distances (up to 10 μm). Our results represent major progress in the quest for a robust, tunable valley pseudospin system among various alternatives [3–5,11,12], and indicate the possibility of using the nonlocal topological transport in practical applications under ambient conditions.

The structure of our dual-gate graphene field-effect transistors (FETs) is shown in Fig. 1a and 1b. We constructed the device by sequentially transferring BLG and hexagonal boron nitride (hBN) flakes onto an hBN substrate supported on a $SiO_2$/Si wafer ($SiO_2$ thickness = 300 nm). Prior to the deposition of the top hBN flake, BLG was etched into a well-defined Hall bar geometry (Fig. 1a, broken line) for easy characterisation in both local and nonlocal configurations. The whole stack was etched again after the deposition of the top hBN to expose the graphene edge for making electrical contacts with metallic leads[13] (Cr/Au, 8 nm and 70 nm, respectively). The final device had BLG sandwiched between the top and bottom gates (Fig. 1b), and the hBN gate dielectric ensured excellent sample quality (see Methods).

Voltages applied on top and bottom gates ($V_t$ and $V_b$) enabled us to independently control the gap opening and carrier doping in the BLG. The bandgap $E_g$ is determined by the average of top and bottom gate-induced electrical displacement fields, $D = (D_t + D_b)/2$, which breaks the inversion symmetry of the BLG[14]. The carrier doping $n$ can be tuned by the difference of the two displacement fields, $n = \varepsilon_0(D_t - D_b)$, where $\varepsilon_0$ is the vacuum permittivity. In our experiment, the drain electrode was grounded, and the displacement fields were related to top and bottom gate voltages by $D_t = \varepsilon_t(V_t - V_{t0})/d_t$ and $D_b = \varepsilon_b(V_b - V_{b0})/d_b$, where $\varepsilon$ and $d$ are the dielectric constant and thickness of the dielectric layers, respectively, and $V_{t0}$ and $V_{b0}$ are effective offset voltages caused by environment-induced carrier doping. The resistance of our sample measured in a standard four-terminal setup (referred to as local resistance $R_L$) exhibited the typical behaviour of BLG: the CNP manifested as a peak in $R_L$ as the carrier density is varied in a pristine sample (Fig. 1c, blue), and the peak value increased substantially as a bandgap was opened by a finite field $D$ (Fig. 2a-d, blue). $R_L$ plotted as a function of both $V_t$ and $V_b$ shows the effect of gate bias more clearly in Fig. 2b. Along the line defined by the $(V_t, V_b)$ pairs at the CNP, the bandgap is fully tuned by $D$, whereas the sample remained charge neutral. The sample exhibited a net charge doping perpendicular to the line in the $(V_t, V_b)$ plane, with the bandgap remaining constant.

A pronounced nonlocal response appeared as the bandgap opens in BLG at low temperatures. We detected the response by sending a current $I$ through the local leads, and sensing the nonlocal voltage $V_{NL}$ at the far end of the sample (see Fig. 1c inset for the measurement setup). The nonlocal resistance $R_{NL}$, defined as $R_{NL} = V_{NL}/I$, is negligible in a pristine sample with zero gap opening (Fig. 1c, orange). A peak in $R_{NL}$, however, appeared at a threshold displacement field of $D = 0.28$ V/nm, and increased rapidly to the order of a few hundred $\Omega$ at large $D$ (Fig. 2a-d, data obtained at temperature $T = 70$ K). The two-dimensional plot of $R_{NL}$ as a function of $V_t$ and $V_b$ shows the general behavior of the nonlocal response at $T = 70$ K. $R_{NL}$ generally peaks at the CNP, but two features distinguish the behavior of $R_{NL}$ from that of the local resistance $R_L$: the $R_{NL}$ peak is generally sharper than the $R_L$ peak, and $R_{NL}$ drops to zero outside of the peak whereas $R_L$ maintains an order of approximately 100 k$\Omega$ at finite doping. Both features indicate the distinct origins of $R_{NL}$ and $R_L$, and are general behavior of nonlocal transport observed in other graphene pseudospin and spin systems[4,7,9]. In the present study, we noted that the peak positions of $R_{NL}$ and $R_L$ were not always aligned with each other; we attribute the misalignment to inhomogeneities that were present in our samples (Supplementary Section IV). We emphasise that the observed $R_{NL}$ was not from the stray charge current that contributes an ohmic nonlocal resistance[7]. Such ohmic contribution decreases exponentially with the sample length-to-width ratio $L/w$, and is up to two orders of magnitude lower than the observed $R_{NL}$ in the device shown in Fig. 2a-d ($L = 5$ μm and $w = 1.5$ μm; the ohmic contribution to $R_{NL}$ is represented by the broken line). A pronounced nonlocal signal was observed in devices with an even higher $L/w$ ratio of up to 6.7 (Fig. 3e).

The temperature dependence of $R_{NL}$ and $R_L$ revealed crucial information on the microscopic mechanism of both the local and nonlocal transport. Three distinct transport regimes were observed at large fields (Fig. 3a): thermal activation at high temperature; nearest-neighbor hopping at intermediate temperature; and variable-range hopping at low temperature. These transport regimes were consistent with previous studies[15,16] (Supplementary Section V). In particular, the high-temperature activation behaviour enabled us to extract the BLG bandgap $E_g$ as a function of $D$, which agreed well with theoretical calculations and previous measurements[14,17] (Fig. 3a inset and Supplementary Section V).

Pronounced nonlocal signal is observed in both thermal activation and hopping regimes (Fig. 3b). We found that $R_{NL}$ also followed an exponential activated behavior in the thermal activation regime at high temperatures (Fig. 3b inset), although with an exponent higher than $E_g/k_B T$ (Supplementary Fig. 5d). The connection between $R_{NL}$ and $R_L$ became obvious when $\ln R_{NL}$ was plotted against $\ln R_L$ (Fig. 3c). Data sets for different $D$ were all in straight lines with the same slope $\alpha = 2.77 \pm 0.02$ in the thermal activation regime; hence, a simple relation $R_{NL} \sim R_L^\alpha$ can be established. We noted that a diffusive nonlocal topological transport model indeed predicted a power law relation $R_{NL} \sim R_L^3 \sigma_{xy}^2 e^{-L/\lambda_v}/\lambda_v$ (ref. 18), where $\sigma_{xy}$ is the valley Hall conductivity, and $\lambda_v$ is the valley diffusion length. Our observation agreed reasonably well with this predicted scaling between $R_{NL}$ and $R_L$, and the deviation of $\alpha$ from 3 may indicate the complicated role of $\sigma_{xy}$ and/or $\lambda_v$ in $R_{NL}$. $\alpha$ varied among samples, probably as a result of differences in sample disorder (Supplementary Fig. 6). Intriguingly, our data show that the simple relation $R_{NL} \sim R_L^\alpha$ persists beyond the thermal activation regime and into the hopping regime until the curves eventually level off (Fig. 3c).

The widely tunable bandgap in BLG provides another crucial benefit, which is the room-temperature operation of our BLG nonlocal FET. Fig. 3d displays the $R_{NL}$ observed up to room temperature in a BLG biased at $D = -1.23$ V/nm (corresponding to $E_g = 135$ meV). A wide bandgap alone does not guarantee high-temperature nonlocal transport because the nonlocal signal, along with $R_L$, decreases exponentially with temperature in the thermal activation regime. However, $R_{NL}$ does not decrease as rapidly with increasing temperature in the hopping regime. When the onset of the hopping regime occurs at high temperatures, room-temperature operation consequently remains possible. The key, therefore, is to find samples in which the onset of the hopping regime occurs at near room temperatures, as we have demonstrated in the samples shown in Fig. 3d and Supplementary Fig. 7.

We now turn to the length dependence of the nonlocal valley transport. Here we note the analog between the valley transport in biased BLG and the spin transport in which spin-flip scattering causes the spin diffusion current to decay exponentially. The valley current in BLG decreases through intervalley scattering, which requires a large momentum transfer (*e.g.*, by atomic scale disorders). Such disorders are found to be extremely rare in cleaved BLG crystals[19,20], implying a large valley diffusion length $\lambda_v$[21,22]. In the present study, an appreciable nonlocal signal was observed in samples up to 10 μm long. Fig. 3e shows the length-dependent $R_{NL}$, measured on a single device under a field of $D = -0.47$ V/nm. From a line fit to the semilog plot of the $R_{NL}$ peak value as a function of sample length (Fig. 3e inset), we obtained an order of magnitude estimation $\lambda_v \sim 1$ μm, because sample inhomogeneity (manifested as shift of $R_{NL}$ peaks in Fig. 3e) preventing a more precise estimation. Such a large length scale agrees reasonably well with recent study on inter-valley scattering[21,22], and is also consistent with the areal density of atomic defects (12.05 μm$^{-2}$) found in Kish graphite[23], which is the same type of specimen used in this study.

The observation of a giant nonlocal response in the middle of the energy gap in insulating BLG was unexpected. Although midgap helical edge sates in 2D quantum spin Hall systems can support long-range nonlocal conduction[24–26], such spin helical edge states did not exist in our study because BLG is topologically trivial. Midgap valley helical modes may still exist at topological domain walls or edges of BLG in certain circumstances[27–29], and may potentially lead to nonlocal conduction. The nonlocal transport through edge states and bulk states, however, exhibit drastically different length dependence; for a given sample width, bulk conduction depends on the active length of the sample, whereas the edge state conduction depends only on the length of the edge[7]. We took advantage of this difference to fabricate the device shown in Fig. 4 inset: two Hall bars (left and right with shared current injection leads 2 and 5) have the same active sample length but substantially different edge lengths. The fact that a comparable nonlocal response was observed on both Hall bars unambiguously demonstrates the bulk origin of our valley transport. A robust nonlocal response with the same sign and order of magnitude was observed in all the samples we fabricated on hBN (6 in total). Helical modes at the domain walls, which differ according to sample if they exist at all, is unlikely to be the origin of the observed nonlocal transport.

The question then arises as to what physical mechanism leads to the nonlocal valley transport that we observed. The valley Hall effect originally proposed in graphene requires finite doping[2], and is therefore not applicable in our insulating BLG. To this end, we note that nonlocal transport in an insulator without helical edge states is an area still open to theoretical investigation[30], and our results call for continued effort, both experimental and theoretical, to address this outstanding problem.

In conclusion, we observed a giant nonlocal valley Hall effect in BLG subjected to a symmetry-breaking gate electric field. The long-distance valley transport is fully tuned by the gate, and persists up to room temperature at large gate biases. Our observation of bulk valley current in biased BLG opens up new avenues for future experimental and theoretical investigation of the valley/spin Hall effect in the insulating regime. In addition, the demonstration of a gate-tunable valley degree of freedom and nonlocal topological transport is major progress towards the development of valleytronic applications.

**Methods**

**Device fabrication.** We fabricated BLG devices by following procedures similar to those described in ref 13 and [31]. BLG flakes transferred onto hBN substrate were etched into Hall bar geometry by using standard electron beam (ebeam) lithography followed by reactive ion etch. The etched BLG Hall bar was then annealed at 370 ℃ in an Ar/H$_2$ atmosphere overnight to remove ebeam resist residue. We further cleaned the sample surface by using atomic force microscope (Park Systems) operating in contact mode. The BLG Hall bar on the hBN substrate was then covered by another layer of hBN as a top-gate dielectric. The hBN/BLG/hBN stack was then etched using inductively coupled plasma (ICP) to expose graphene leads at the edge. Finally, metal electrodes (Cr/Au, 8 nm and 70 nm) were deposited for electrical contact with graphene leads. The devices exhibited mobility ranging from 1,500 cm$^2$/Vs to 50,000 cm$^2$/Vs at 10 K, and nonlocal signal was observed in all of them.

**Nonlocal resistance measurement.** The high common-mode voltage present in the measurement circuit and the high output impedance of the nonlocal voltage signal cause the typical lock-in measurement to fail at large gate biases when BLG becomes excessively resistive. We found that the problem can be mitigated by increasing the input impedance of the preamplifier of the measurement circuit. Reliable measurement of the nonlocal signal can also be obtained by detecting the current instead of voltage in the nonlocal leads; however, precautions must be taken to float the current preamplifier. A detailed discussion of our nonlocal measurement is presented in Supplementary Section I.

**Acknowledgements**

We thank Q. Niu, D.-H. Lee, F. Wang, J. Shi, J. Xiao, J. Zhu for the discussions. Part of the sample fabrication was performed at the Fudan Nano-fabrication Lab. M.S., G.C., L.M. and Y.Z. acknowledge the financial support of the National Basic Research Program of China (973 Program) under the grant nos. 2013CB921902 and 2011CB921802, and NSF of China under the grant nos. 11034001 and 11425415. W.Y. acknowledges the support from the University of Hong Kong (OYRA), and the RGC of Hong Kong SAR (HKU706412P). W.S. and D.X. acknowledge the support from the U.S. Department of Energy, Office of Science, Office of Basic Energy Science, under award no. DE-SC0012509.

**Figure captions**

**Figure 1 | Dual-gated BLG FET and its local and nonlocal characterisation. a**, Optical image of a typical BLG FET viewed from the top. A BLG Hall bar (outlined by a broken line) is sandwiched between the top and bottom hBN flakes. The hBN/BLG/hBN stack (blue) was etched to expose the BLG leads at the edge which are subsequently contacted by Au electrodes (orange). **b**, Schematic cross-sectional view of the device shown in **a**. An Au pad and degenerately doped silicon served as top and bottom gate, respectively. **c**, Local and nonlocal resistance measured as a function of top-gate voltage $V_t$ while the bottom-gate voltage $V_b$ was fixed at zero. BLG should have a zero bandgap at the CNP, and nonlocal signal was not detected under this condition. Inset: Illustration of the local and nonlocal measurement setup.

**Figure 2 | Local and nonlocal response of biased BLG. a-d**, Local and nonlocal resistance as a function of $V_t$ with $V_b$ fixed at varying voltages. The length and width of the sample were 5 μm and 1.5 μm, respectively. The broken lines show the expected ohmic nonlocal contribution (see text), which is substantially smaller than the actual measured $R_{NL}$ (orange curves). **e** and **f**, local and nonlocal resistance measured as a function of both $V_t$ and $V_b$. Data were obtained from the same device measured in **a-d**. The crosses mark the point $(V_{t0}, V_{b0}) = (1.3V, -7.5V)$ where no bandgap opened in BLG. All data were recorded at $T = 70$ K.

**Figure 3 | Temperature and length dependence of the local and nonlocal responses of biased BLG a**, Peak resistance in $R_L$ measured as a function of temperature under varying electric displacement fields. Three transport regimes are clearly visible for displacement field $D > 0.5$V/nm (see text). Inset: The Arrhenius plot of $R_L$ showing thermal activation behavior at high temperatures. Broken lines are the line fit to the linear part of the Arrhenius plot, and thermal activation energy could be obtained from the slope of the line fit (Supplementary Section V). The triangles mark the onset of hoping regime. **b**, Peak nonlocal resistance in $R_{NL}$ measured as a function of temperature under varying electric displacement fields. Data were obtained from the same device measured in **a**. Inset: The Arrhenius plot of $R_{NL}$. $R_{NL}$ also shows thermal activation behavior at high temperatures, and broken lines are the line fit. **c**, $\ln R_L$ plotted against $\ln R_{NL}$ for varying displacement field $D$. $R_{NL}$ and $R_L$ are from the data sets in **a** and **b**. Triangles mark the onset of hoping regime shown in **a**. $\ln R_L$ scales linearly with $\ln R_{NL}$ with a single slope $\alpha = 2.77 \pm 0.02$ for different $D$ in both thermal activation and hopping regimes, before leveling off deep in the hopping regime. The broken lines are the line fit to the linear part of the data sets. **d**, High-temperature

$R_{NL}$ obtained in a sample in which the onset of hopping regime occurred at near room-temperature (see text). The curves are shifted for clarity. The broken lines are a visual guide. The back gate was fixed at $V_b = -100$ V. **e**, Length dependence of nonlocal response. Hall bars with different length $L$ were fabricated on a single a device. $R_{NL}$ was measured at different $L$ with $V_b = 30$ V and T = 20 K. Sample width: 1.5 μm. Inset: Semilog plot of peak values of $R_{NL}$ as a function of $L$. The broken line is a visual guide that corresponds to a valley diffusion length of approximately 1 μm.

**Figure 4 | Bulk v.s. edge nonlocal transport.** Nonlocal resistance as a function of $V_t$ ($V_b$ fixed at 50 V) obtained on two Hall bars (left and right) shown in the inset. Data were obtained at T = 30 K. The two Hall bars share two local leads (2 and 5) in the middle, and the left and right Hall bar have the same active sample length but substantially different edge lengths. We passed current through the two local leads and measure $R_{NL}$ at the far end of the left Hall bar (contact 1 and 4) and right Hall bar (contact 3 and 6). Nonlocal signal was detected with similar amplitude on both the left and right Hall bars, indicating the bulk origin of the nonlocal transport. Inset: BLG sample on an hBN substrate before deposition of the top hBN flake. The purple dashed line outlines the geometry of etched BLG, and the yellow dashed lines indicate the position of the electrodes to be deposited.

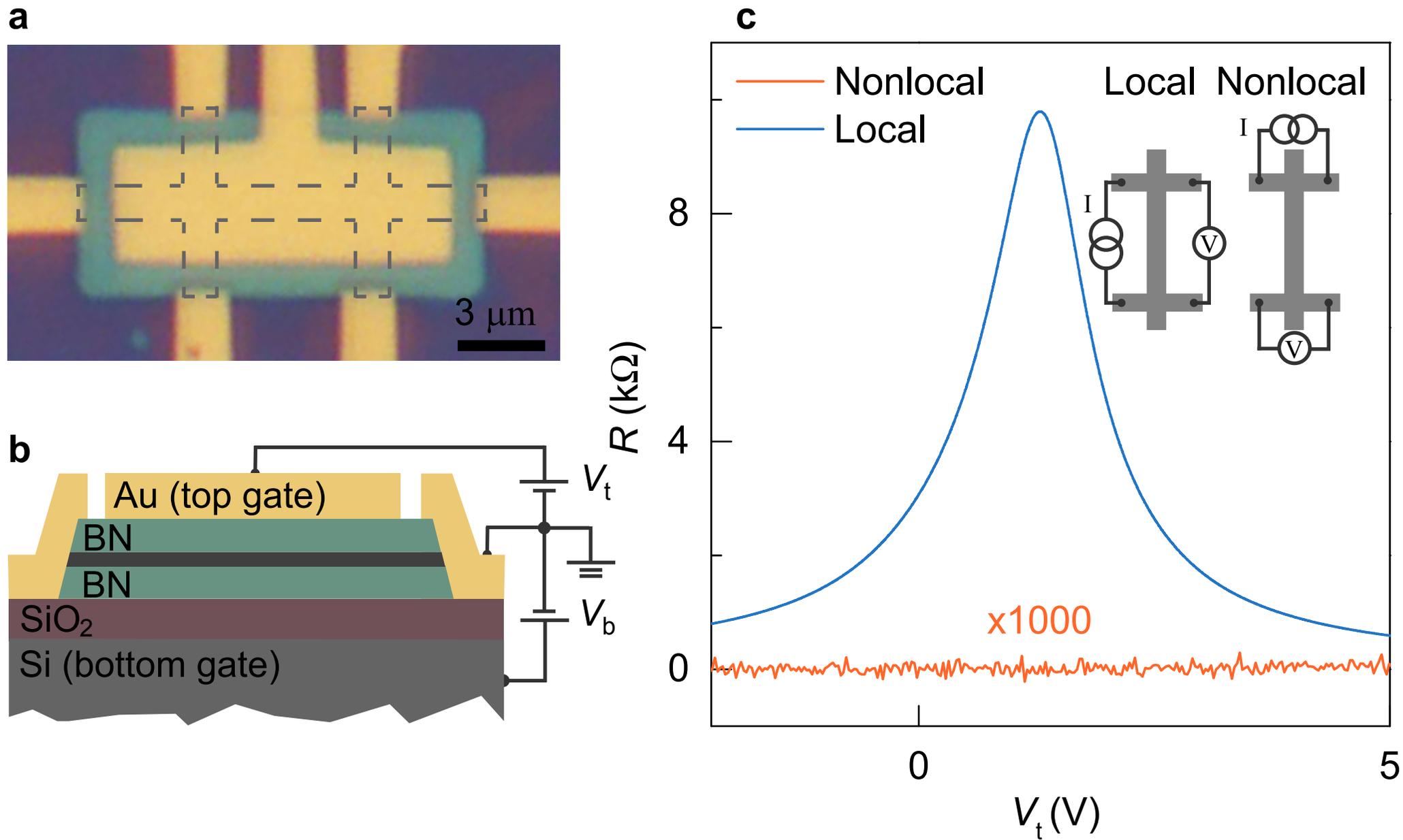

Fig. 1, M. Sui *et al*

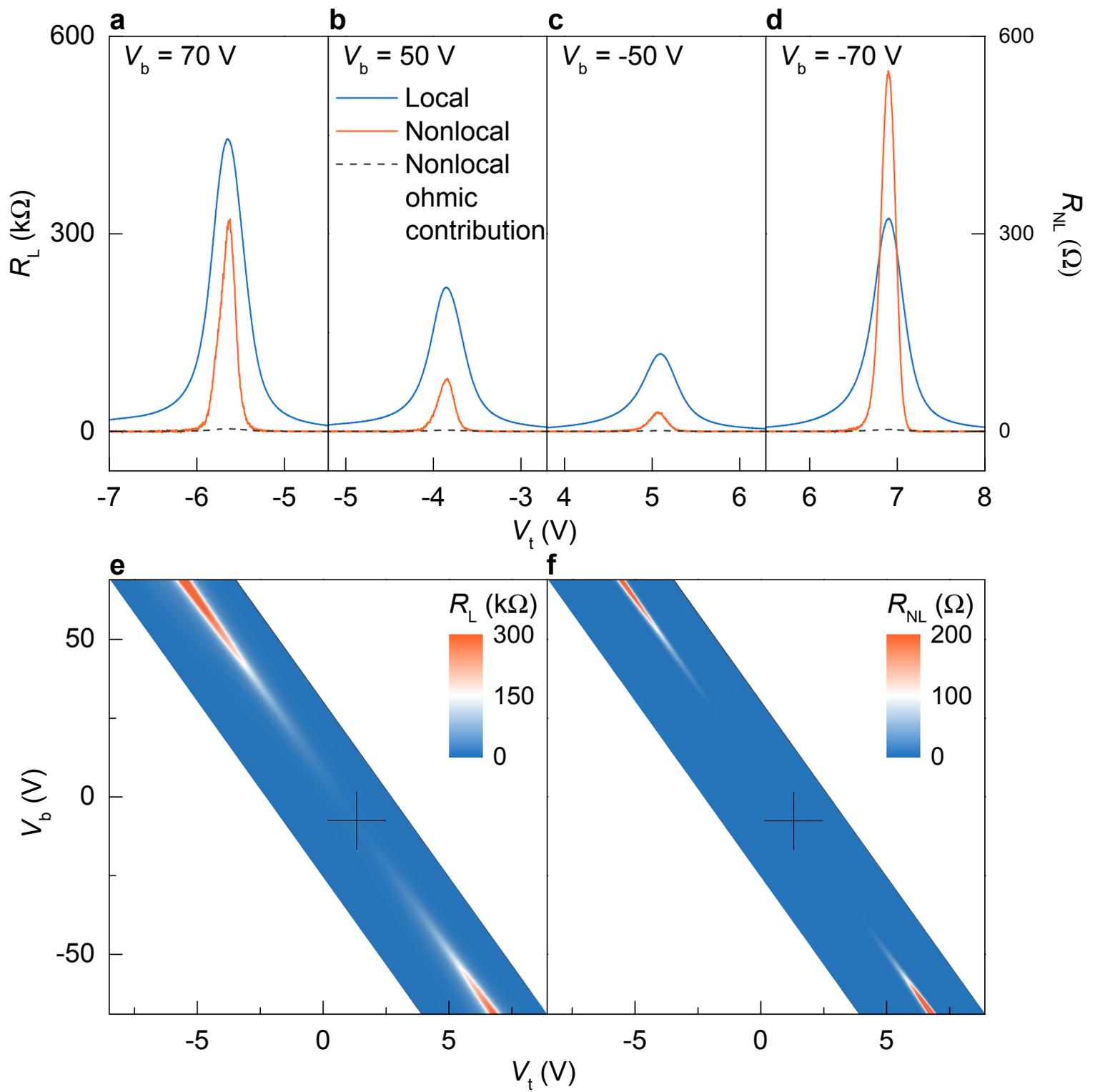

Fig. 2, M. Sui *et al*

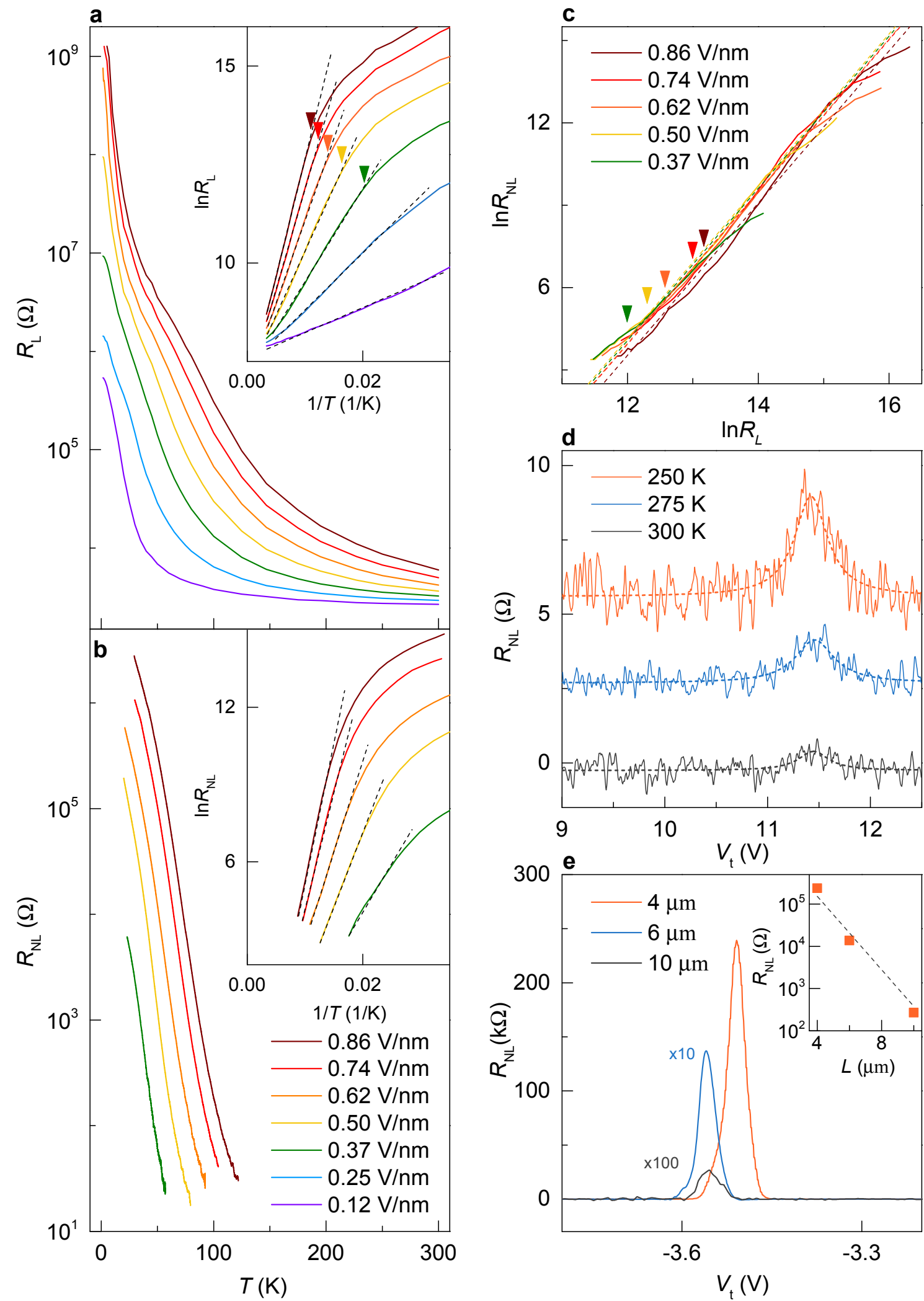

Fig. 3, M. Sui et al

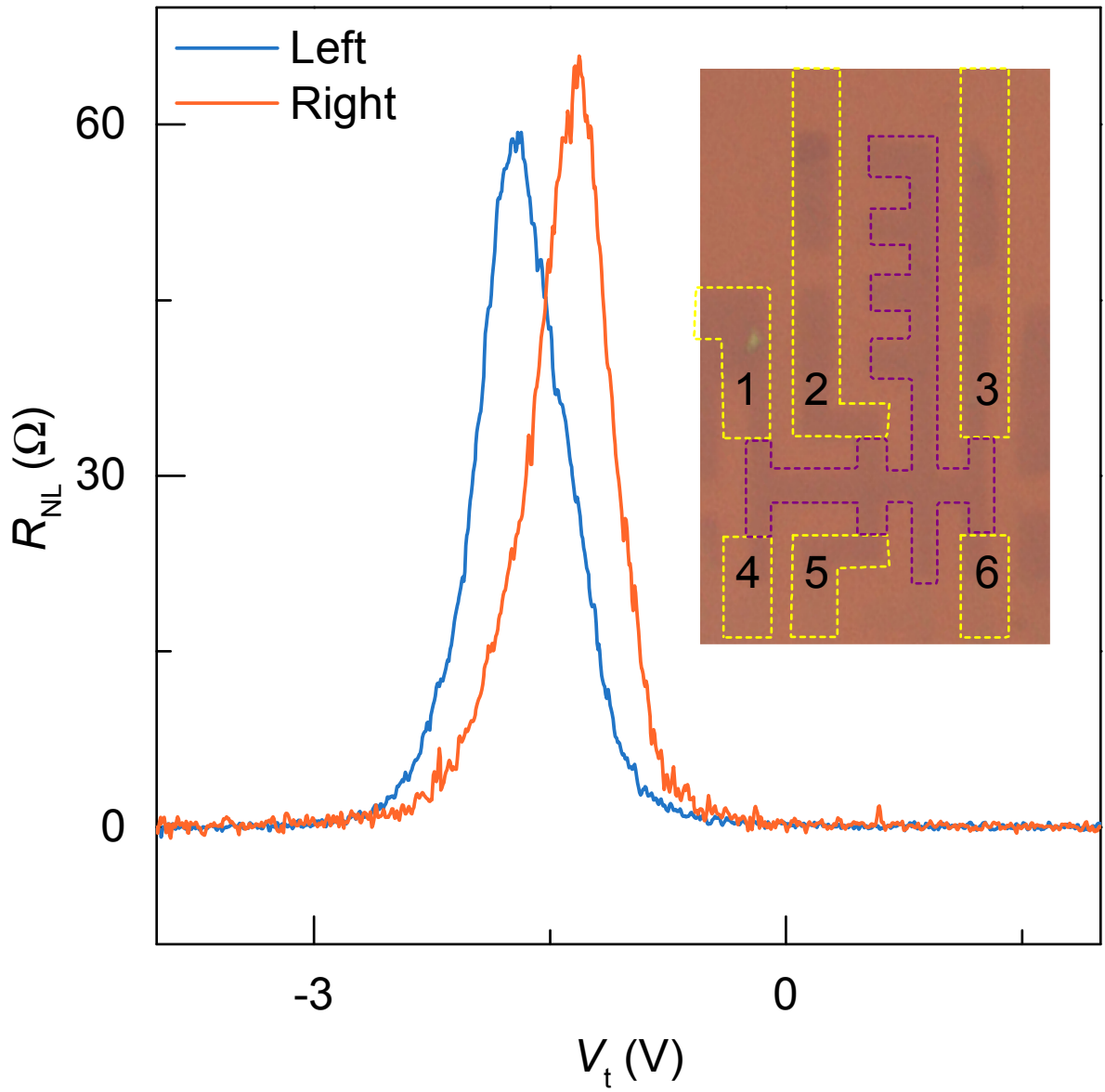

Fig. 4, M. Sui *et al*

# Supplementary Information for

# Gate-tunable Topological Valley Transport in Bilayer Graphene

Mengqiao Sui, Guorui Chen, Liguo Ma, Wenyu Shan, Dai Tian, Kenji Watanabe,

Takashi Taniguchi, Xiaofeng Jin, Wang Yao, Di Xiao and Yuanbo Zhang[*]

[*]Email: zhyb@fudan.edu.cn

## Content

# I. Measurement schemes to eliminate artifacts in nonlocal detection

Measurement artifacts may appear when typical lock-in measurement setup (shown in Fig. S2b) is used for detection. An example of the artifact is shown in Fig. S1a, where the peak in $R_{NL}$ are distorted and negative $R_{NL}$ appears. We found that the spurious signal was from the measurement setup, as we shall discuss below.

The spurious signal came from the common mode voltage $V_{CM}$ at the current injection side of the Hall bar (marked by the red dot in Fig. S1b). Since lock-in amplifier input typically has a single-ended input resistance of $R_{input} \sim 10$ MΩ to the ground (we use Stanford Research Systems SR830), $V_{CM}$ induces current through the two nonlocal leads (3 and 4 in Fig. S1b). Resistance imbalance between the two nonlocal leads could therefore induce a spurious voltage difference at the input of the lock-in amplifier:

$$V_{NL}^s = \frac{V_{CM} R_A}{R_{input} + R_A} - \frac{V_{CM} R_B}{R_{input} + R_B} \approx (R_A - R_B) \frac{V_{CM}}{R_{input}} \qquad (1)$$

where $R_A$ and $R_B$ are the total resistance from cross at the local side of the Hall bar (red dot in Fig. S1b) to the input of the lock-in amplifier. $R_A$ and $R_B$ include BG resistance, contact resistance at the nonlocal electrodes (3 and 4 in Fig. S1b), and resistance of the voltage leads, and are typically on the order of 1 MΩ. In a typical measurement, $V_{CM} \approx 10$ mV assuming a sample resistance on order of 1 MΩ and current excitation of 10 nA. A difference of $\sim 10$ kΩ in $R_A$ and $R_B$, which is also typical, would introduce a $V_{NL}^s$ on the order of $\sim 1$ μV. Such a spurious voltage is on the same order of magnitude as the real valley Hall nonlocal signal, and is indeed what we have measured in Fig. S1a. We note that incomplete rejection of $V_{CM}$ at the lock-in pre-amplifier, as well as the capacitive coupling between the voltage probe leads, could also induce spurious effects in principle. Those effects, however, can be suppressed 1) by the high common mode rejection ratio $\sim 10^5$ found in typical lock-ins, and 2) by operating the lock-in at low frequencies ($\sim 3$ Hz in our case).

The spurious voltage $V_{NL}^s$ can be eliminated by careful design of the measurement setup. Here we describe three measurement schemes that are able to detect true nonlocal valley Hall signal without distortion.

First, $V_{NL}^s$ can be reduced by increasing the input impedance of the voltage meter, as is apparent in Eq. (1). This can be achieved by replacing lock-in amplifier with high input impedance voltage meters such as Keithley 2182A nanovoltmeter ($R_{input} > 10$ GΩ) as shown in Fig. S1c. We found that the nanovoltmeter, used in combination with Keithley 6221 precision current source operating in delta mode, yielded satisfactory measurement results as shown in Fig. S1f (Method A).

Second, $V_{NL}^s$ can also be reduced by suppressing $V_{CM}$. To this end, we took advantage of a feedback circuit consisting of an operational amplifier (OPA), and devised a measurement setup schematically shown in Fig. S1d. The OPA kept the voltage level at the cross center (marked by blue dot in Fig. S1d) at zero while maintaining the same current flow, so the measurement circuit was not disturbed. We confirmed that this method suppresses $V_{CM}$ down to $< 1$ μV, and gave a good measurement of $R_{NL}$ as shown in Fig. S1f (Method B).

Third, the effect of $V_{CM}$ can be eliminated altogether by floating the nonlocal signal detection setup, and measuring the nonlocal current instead of voltage as illustrated in (Fig. S1e). To float the current amplifier (DL Instruments Ithaco Model 1211), we powered it with battery, and used an optical isolator to retrieve the output signal without bridging the ground. The nonlocal voltage (Fig. S1f, Method C) was then obtained by multiplying the measured nonlocal current with the internal impedance of the nonlocal signal.

The three independent measurement gave consistent nonlocal measurement results (Fig. S1f). The excellent agreement among them indicated that the spurious voltage $V_{NL}^s$ can be effectively suppressed.

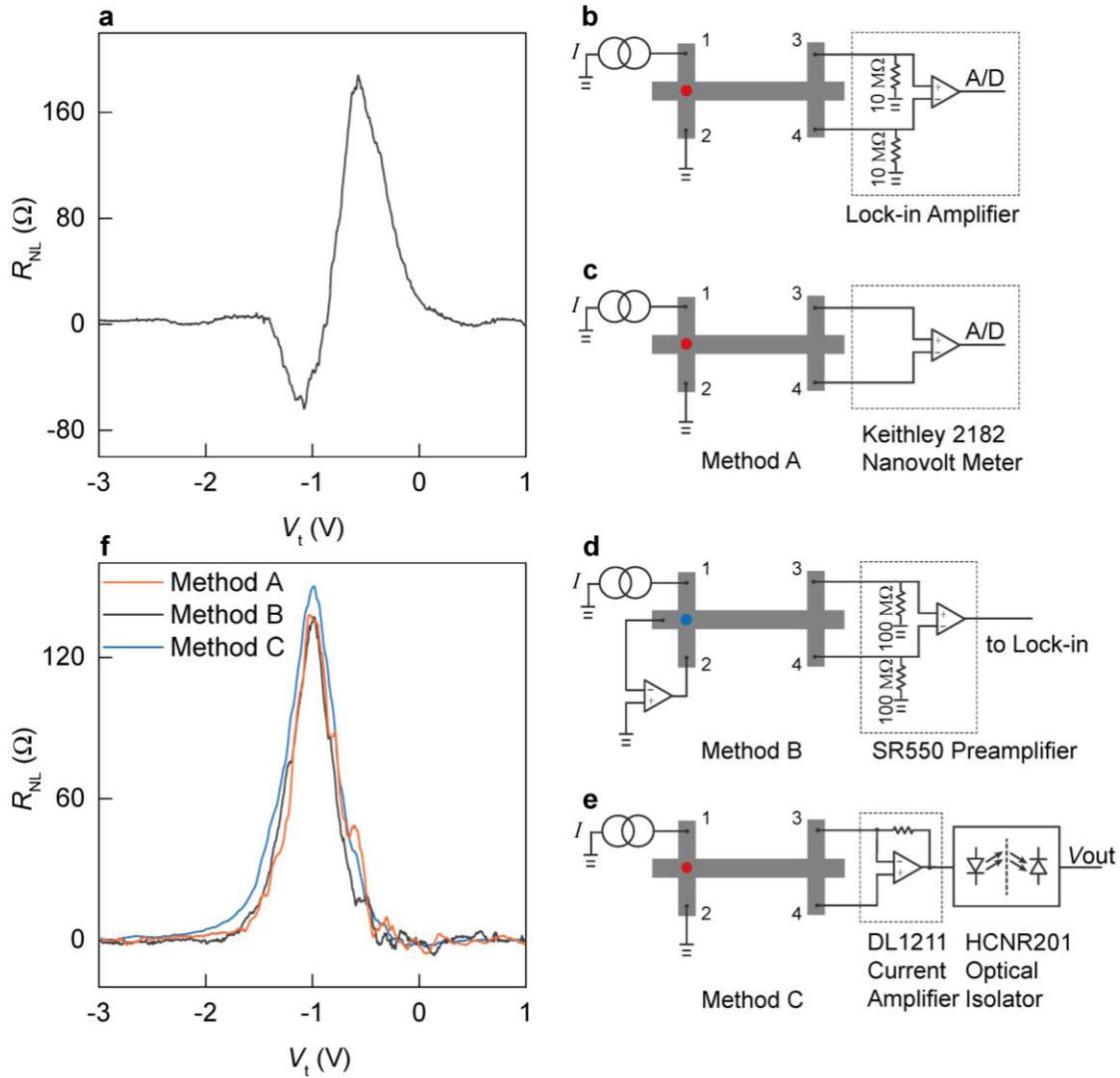

**Supplementary Figure 1 | Nonlocal measurement schemes to eliminate spurious signal. a**, Nonlocal resistance $R_{NL}$ as a function of top gate voltage $V_t$ obtained from typical lock-in measurement. The valley Hall signal mixed with spurious signal to produce the curve shown here. **b**, Nonlocal voltage measurement with typical lock-in amplifier (typical input impedance of $R_{input} \sim 10\ \text{M}\Omega$). **c**, Nonlocal voltage measurement with Keithley 2182A nanovoltmeter ($R_{input} > 10\ \text{G}\Omega$). **d**, Nonlocal voltage measurement with voltage preamplifier (Stanford Research Systems, Model SR550), while the common mode voltage $V_{NL}^s$ is suppressed by a voltage-balancing OPA. **e**, Nonlocal current measurement with a floated current amplifier. The output of the amplifier is transmitted through an optical isolator to keep the amplifier floated. **f**, $R_{NL}$ as a function of $V_t$ obtained using the methods described in **c**, **d** and **e**. The sample is the same as in **a**, but the spurious signal is now eliminated from the measurements. The results from three independent measurements agree with each other well.

## II. Characterizing nonlocal voltage signal

We provide basic characterization of the nonlocal voltage signal in this section. To ensure reliable nonlocal voltage measurement, we made sure that the electrical contacts to BG was ohmic. This is shown in Fig. S2a, where the linear I-V characteristics at the CNP of biased BG indicate that the contacts remain ohmic even when BG is gapped. The measurement was done at $T = 70$ K, the same temperature where our nonlocal measurements in Fig. 2 were performed.

We then performed a set of basic checks on the nonlocal voltage measurements to rule out artifacts potentially from other sources such as heating effect[1]. The nonlocal voltage changed its sign when the local current injection was reversed (Fig. S2b), indicating that heating effect was negligible in our setup. We also checked the 2f output in our lock-in measurement, did not observe discernable signal from heating effect. In addition, our nonlocal measurement satisfied the universal reciprocal relation, i.e. we observed the same signal if the current and voltage probes were switched (Fig. S2b).

Finally, we characterized the output impedance of the nonlocal voltage. In the measurement setup shown in Fig. S2c, nonlocal current is a function of a known external resistance $R_0$ we add in the circuit:

$$I_{NL} = \frac{IR_{NL}}{R_0 + R_s} \qquad (2)$$

where $R_s$ represents the output impedance of the nonlocal voltage signal (with lead resistance neglected). Eq. (2) implies a linear relation between $R_0$ and $I/I_{NL}$: $R_0 = R_{NL}(I/I_{NL}) - R_s$. Indeed, as we changed $R_0$ from 0 to 2 MΩ, $R_0$ v.s. $I/I_{NL}$ fell on a straight line, whose slope and intercept yielded a measurement of $R_{NL}$ and $R_s$, respectively (Fig. S2d). We noted that the $R_{NL}$ obtained this way agreed with that from independent measurements described in Section I under the same condition. The $R_s$ of 0.5 MΩ also agreed with the sample resistance between the nonlocal probes determined separately. The fact that $R_{NL}$ and $R_s$ were independent of excitation current (Fig. S2d) corroborate our finding that no sample heating occurred during the measurement.

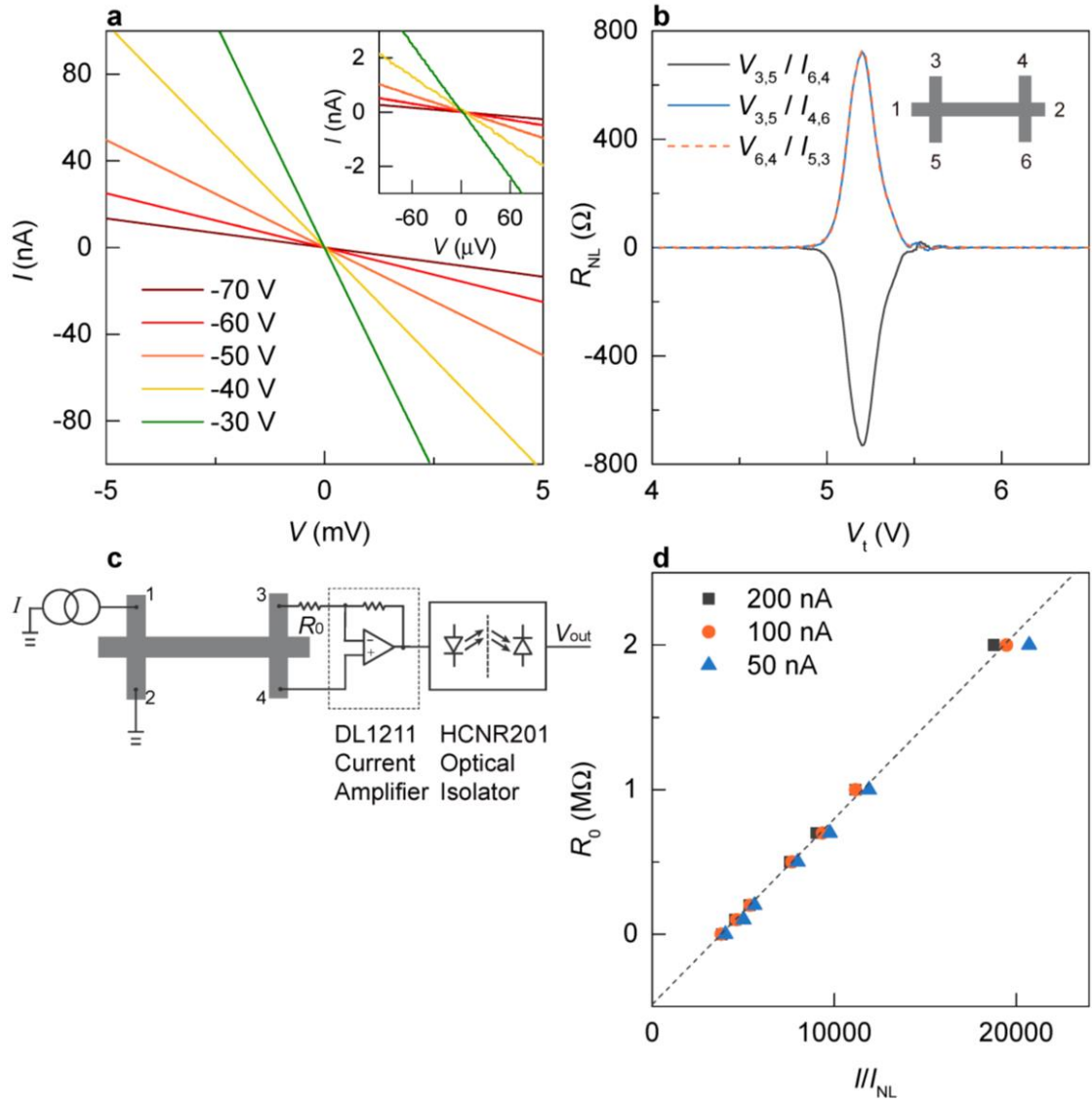

**Supplementary Figure 2 | Nonlocal voltage signal characterization. a**, Two-terminal I-V characteristics at CNP of a BG under varying gate-biases. Inset: zoomed in view of the I-V characteristics at low-biases. Linear I-V indicates ohmic contacts. Data were taken from the same sample and under the same condition as in Figure 2 in the main text. $T = 70$ K. **b**, Nonlocal signal measurement with reversed current (black and blue), and with voltage and current leads swapped (blue and orange). Measurement was performed at $T = 30$ K with $V_b$ kept at $-50$ V. **c,** Nonlocal current measurement circuit with an external resistor $R_0$ added in series with the nonlocal voltage output. **d**, $R_0$ plotted as a function of $I/I_{NL}$ for excitation current of 200 nA (black), 100 nA (orange) and 50 nA (blue). The broken line is a linear fit to the data.

## III. Additional sets of nonlocal measurement

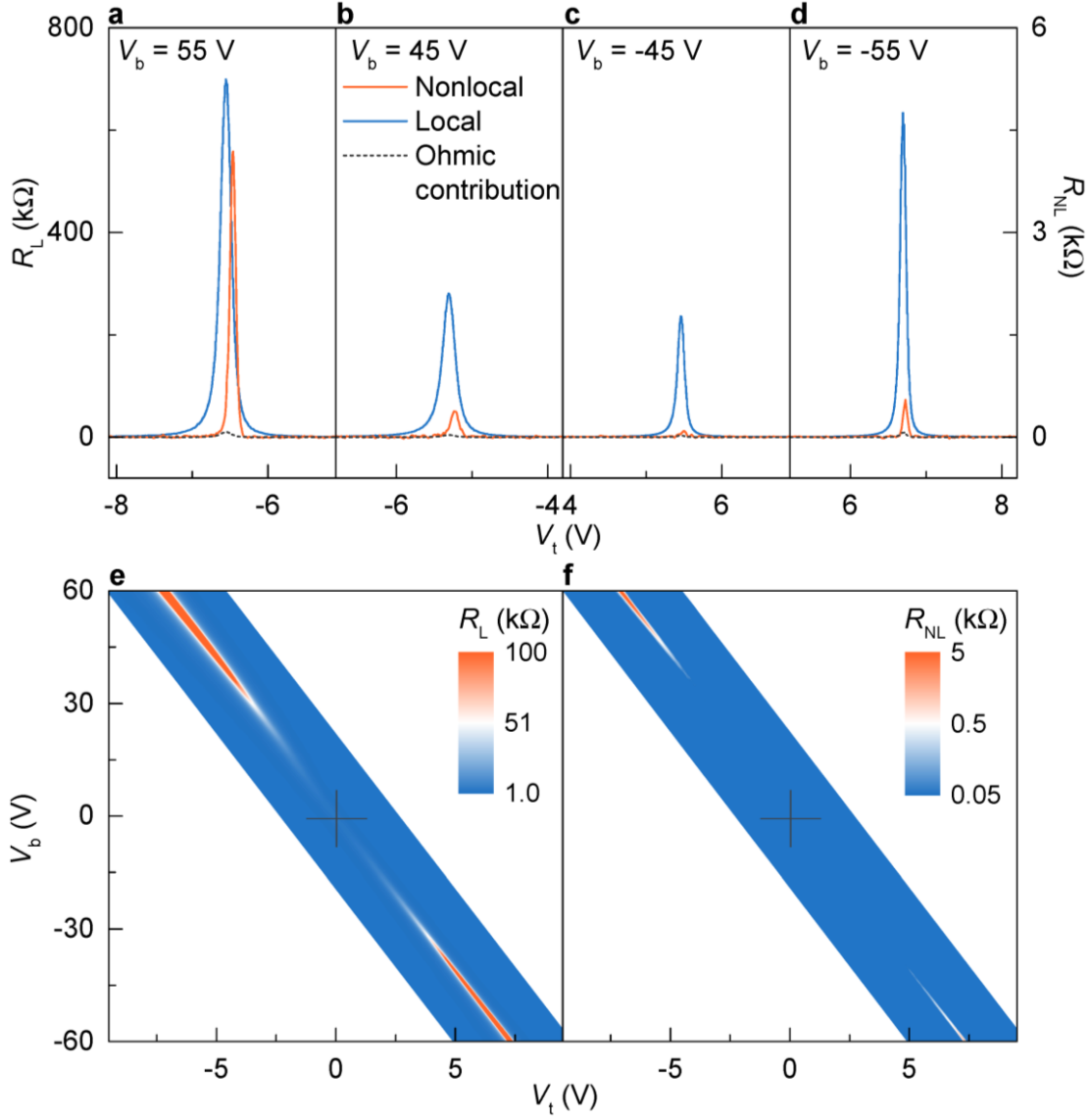

**Supplementary Figure 3 | Local and nonlocal response of a biased BG different from the one shown in Fig. 2. a-d**, Local and nonlocal resistance as a function of $V_t$ with $V_b$ fixed at different values. The length and width of the sample are 4 μm and 1.5 μm, respectively. Broken lines show expected ohmic nonlocal contribution, which is again much smaller than the observed $R_{NL}$ (orange). We note that the peaks in $R_{NL}$ are not aligned with peaks in $R_L$. Such misalignment was sample dependent, and we found it most likely due to inhomogeneity present in some of our BG samples (see section IV). **e** and **f**, local and nonlocal resistance measured as a function of both $V_t$ and $V_b$. Data were taken on the same device measured in **a-d**. The crosses mark the point ($V_{t0}$, $V_{b0}$) = (0.035 V, −0.7 V), where no bandgap is opened in BG. All data were recorded at $T = 40$ K from the same sample shown in Fig. 3 a-c in the main text.

## IV. Role of sample inhomogeneity

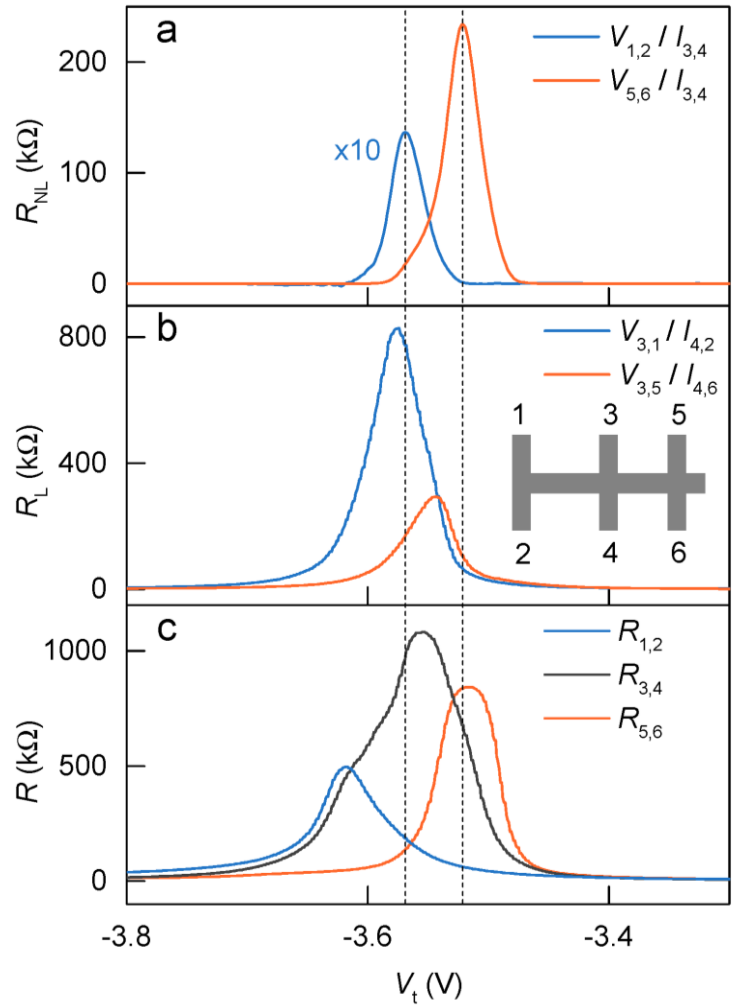

**Supplementary Figure 4 | Sample inhomogeneity manifested as shift of resistance peaks at CNP. a** and **b**, Nonlocal and local resistance, respectively, of the same sample measured in Fig. 3e. **c**, Two-terminal resistance $R$ of the same sample as in **a** and **b**. A schematic of the device is shown in the inset of **b**. The broken lines mark the nonlocal peak positions. The different peak position in $R_L$ and $R$ obtained from different parts of the sample is a result of sample inhomogeneity[2]. The peaks in $R_{NL}$ appear at positions close to peaks in $R_L$ and $R$ measured in similar part of the sample, pointing to sample inhomogeneity as the common origin of the peak position variations in $R_{NL}$, $R_L$ and $R$.

# V. Thermally activated and hoping transport in bilayer graphene

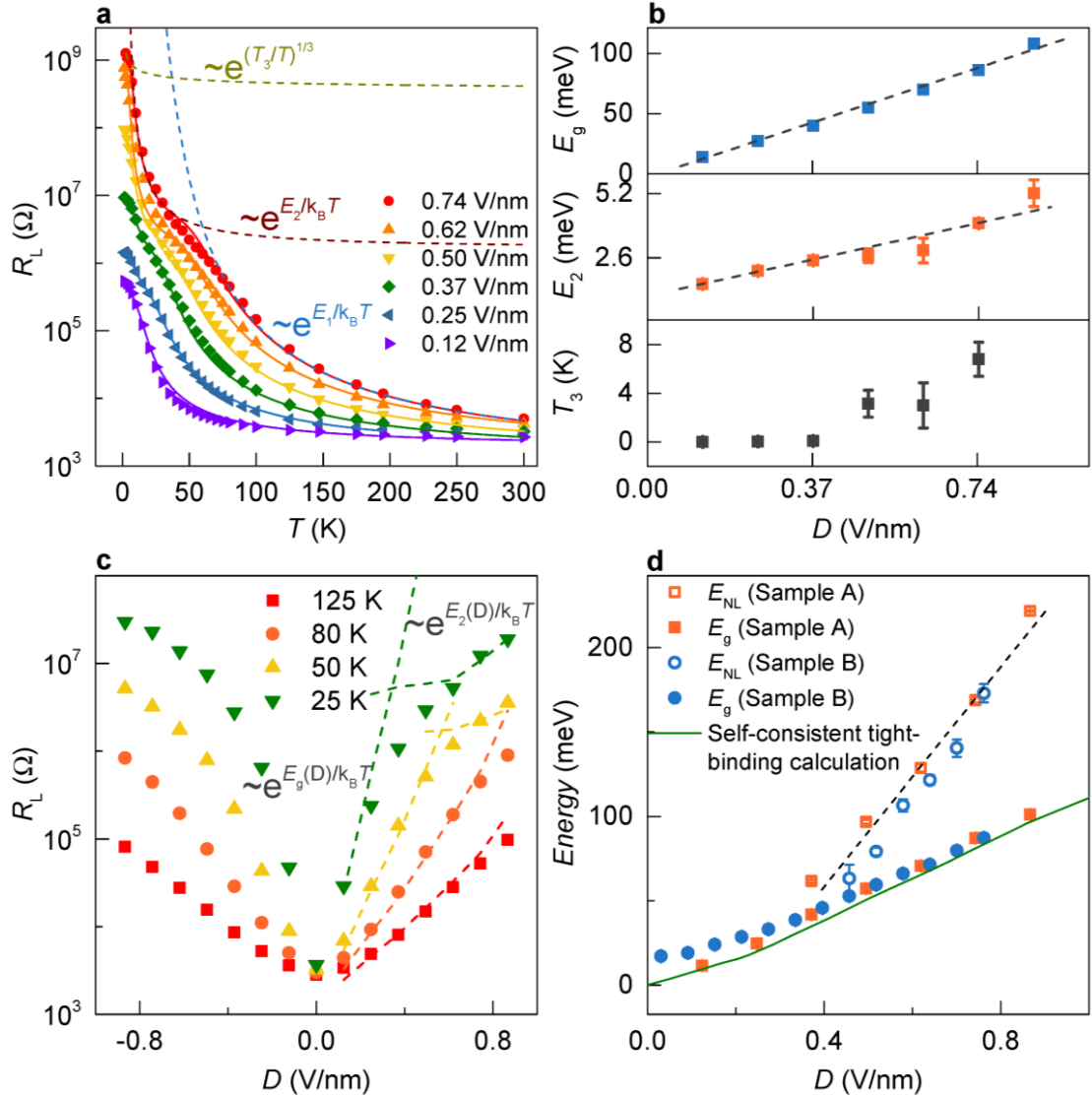

**Supplementary Figure 5 | Temperature-dependent local and nonlocal transport in BLG. a**, $R_L$ as a function of temperature at varying electric displacement fields. The same set of data was shown in Fig. 3a in the main text. The behaviour of the observed $R_L$ agreed well with that reported in ref 3–5, and all the data sets can be well fitted with the relation $R_L^{-1} = R_1^{-1} exp(-E_g/2k_BT) + R_2^{-1} exp(-E_2/k_BT) + R_3^{-1} exp[-(T_3/T)^{1/3}]$, where the three terms correspond to contributions from thermal activation, nearest-neighbor hopping and variable-range hopping, respectively[4] (solid lines are the fitting results). We obtained energy scales $E_g$, $E_2$ and $k_BT_3$ from the fitting, and $R_1$, $R_2$ and $R_3$ are free parameters. $E_g$ gives a measurement of BLG band gap; $E_2$ is the nearest-neighbor hopping energy; and $T_3$ is the onset temperature of variable-range hopping conduction. The broken lines represents individual contributions from thermal activation, nearest-neighbor hopping and variable-range hopping obtained from the fitting. **b**, $E_g$, $E_2$ and $k_BT_3$ as functions of $D$ obtained from the fitting in **a**. Both $E_g$ and $E_2$ increase linearly with $D$, and the broken lines are guide

to the eye. Finite contribution from variable-range hopping appeared at $D \geq 0.5$ V/nm. These results are consistent with those reported in ref 3–5. **c**, Peak values of $R_L$ at CNP measured as a function of $D$. At low temperatures (25 K and 50 K), the transport is dominated by thermal activation and nearest-neighbour hoping at low $D$ and high $D$, respectively. Both thermal activation and nearest-neighbour hoping give exponential dependence of $R_L$ on $D$, and the broken lines are plots of the two contributions with parameters obtained from fitting in **a**. At high temperatures (T $\geq$ 80 K), thermal activation alone dominated the transport at all $D$. **d**, $E_g$ (solid symbols) and the exponent $E_{NL}$ of nonlocal transport in the thermal activation regime (empty symbols) plotted as a function of $D$. Data were obtained by fitting of the slope at the linear part in the inset of Fig. 3a and 3b. The data were obtained from two samples: sample A is the same sample shown in Fig. 3a-c; sample B is the same sample shown in Fig. 2. The measured $E_g$ agrees well with the BLG bandgap calculated with self-consistent tight-binding method (solid green line). The broken line is a linear fit to $E_{NL}$. $E_{NL}$ is much larger than $E_g$, probably because the power law relation between $R_{NL}$ and $R_L$: $R_{NL} \sim R_L^\alpha$. Indeed, the slope of $E_{NL}$ is ~2.8 times larger than that of $E_L$, which is consistent with $\alpha = 2.77 \pm 0.02$ obtained from the scaling between $R_{NL}$ and $R_L$ (see main text).

# VI. Sample dependence of the scaling between $R_{NL}$ and $R_L$

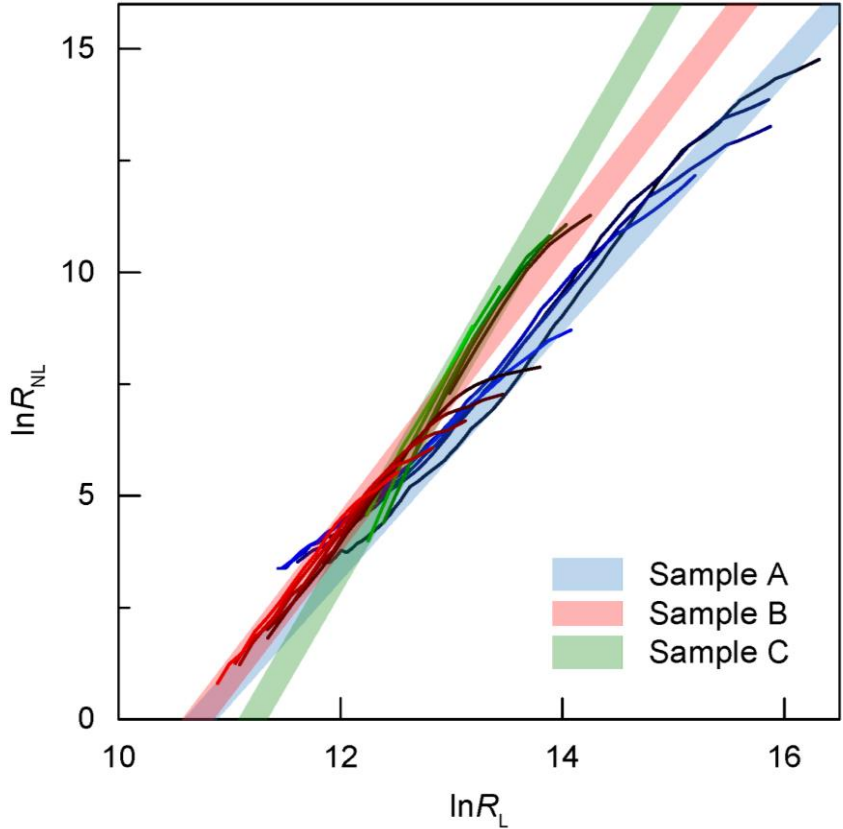

**Supplementary Figure 6 | Sample-dependent scaling between $R_{NL}$ and $R_L$.** $\ln R_L$ plotted against $\ln R_{NL}$ at varying displacement field $D$. Date from three different samples are shown, with sample A the sample shown in Fig. 3c. The displacement fields are: $D = 0.37$ V/nm to $0.86$ V/nm in $0.12$ V/nm steps (red to black) for sample A; $D = 0.52$ V/nm to $0.76$ V/nm in $0.061$ V/nm steps (blue to black) for sample B; $D = 0.35$ V/nm to $0.83$ V/nm in $0.12$ V/nm steps (green to black) for sample C. The power law relation $R_{NL} \sim R_L^\alpha$ described all three data sets with the fitted value of $\alpha$ at 2.77, 3.26 and 4.27 in sample A, B and C, respectively. The curves all leveled off deep in the hopping regime, and those part of the curves was not included in the fitting.

# VII. High-temperatures local and nonlocal transport in different samples

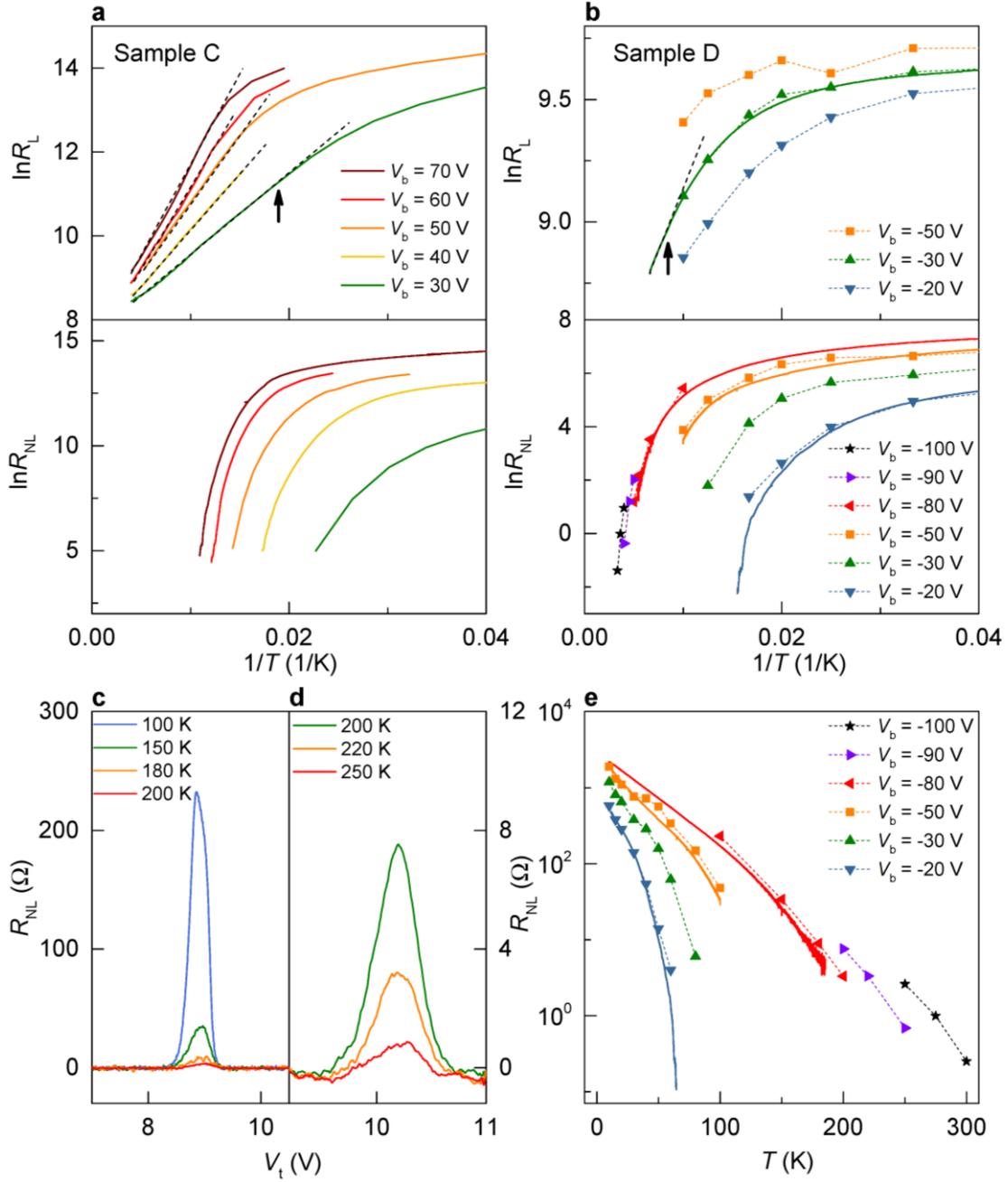

**Supplementary Figure 7 | Thermally activated transport at high temperatures in different samples. a** and **b**, Arrhenius plots of peak resistance of $R_{NL}$ and $R_L$ at CNP as functions $1/T$ for sample C and D (sample D is the same sample shown in Fig. 3d). The solid lines were obtained by sweeping temperature at CNP, and the scattered data points were obtained from gate sweeps at fixed temperatures. The black broken lines are linear fit in the thermal activation regime. The onset of the hopping regime (black arrows) is different in the two samples under comparable electric displacement fields. In both samples, $R_{NL}$ decreased slowly with increasing temperature in the hopping regime, and dropped rapidly to 0 in the thermal activation regime. Such behaviour is

consistent with the substantially different energy scale $E_g$ and $E_2$ (Supplementary Section V). High temperature operation is therefore possible in samples with high onset temperature. **c,** High temperature $R_{NL}$ measurement in sample C at $V_b = -80$ V. Finite $R_{NL}$ persisted up to 200 K. **d,** High temperature $R_{NL}$ measurement in sample C at $V_b = -90$ V. Finite $R_{NL}$ was observed up to 250 K. **e,** Peak $R_{NL}$ at CNP as a function of temperature obtained from sample D. Finite $R_{NL}$ persisted up to 300K at $V_b = -100$ V.

## VIII. Absence of nonlocal signal in monolayer graphene

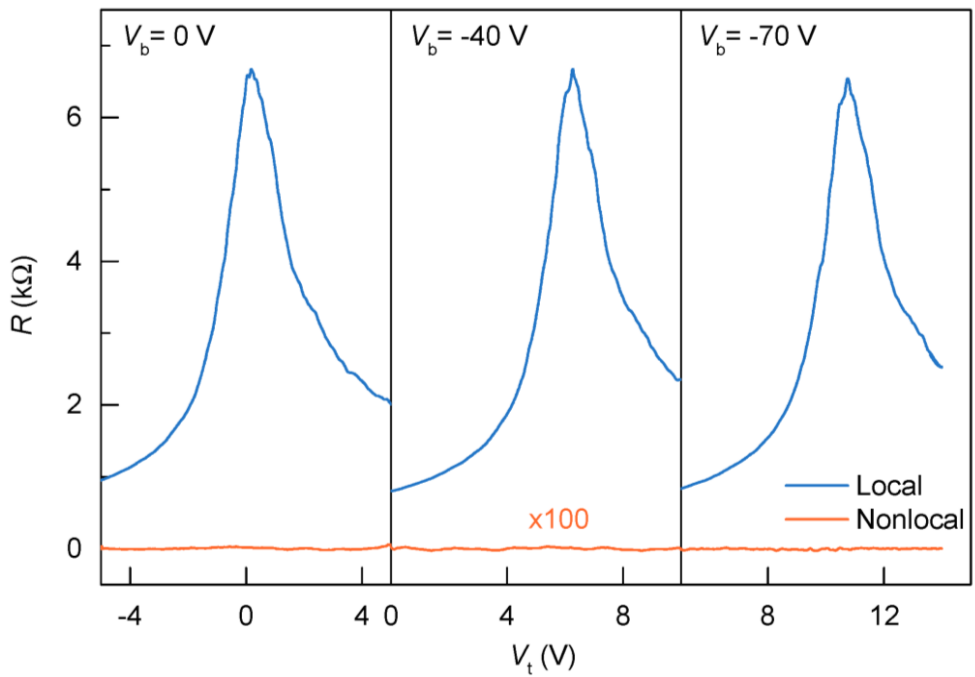

**Supplementary Figure 8 | Local and Nonlocal measurement in monolayer graphene.** Monolayer graphene FET with dual-gate structure was fabricated with the same procedure discussed in Methods. The length-to-width ratio is L/w = 5, and the mobility is approximately 5000 cm$^2$V$^{-1}$s$^{-1}$ at 10 K. The local and nonlocal resistance (blue and orange, respectively) were recorded at $V_b = 0$ V, $-40$ V and $-70$ V, covering the typical range of gate electric field that we used in BLG measurements. The peak value of local resistance did not change with gate, and we observed zero nonlocal response, to the best of our measurement resolution ($< 1$ Ω ), for all the applied gate electric fields. The absence of nonlocal response in gate-biased monolayer graphene highlight the crucial role of gate-induced symmetry breaking in the nonlocal transport in BLG.